\documentclass[11pt]{article} 
\usepackage{hyperref} 
\usepackage{graphicx}

\begin{document} 
 
\title{A fountain of droplets} 
 
\author{D. Terwagne, G. Delon, N. Adami, N. Vandewalle, H. Caps, and S. Dorbolo\\
\\\vspace{6pt} GRASP, D\'epartement de Physique, Universit\'e de Li\`ege \\
B-4000 Li\`ege, Belgium } 
 
\maketitle

\begin{abstract} 
A vessel is plunged upside down into a pool of 50 cSt silicone oil. An air bell is then created. This bell is vertically shaken at 60 Hz that leads to the oscillation of the air/oil interface. The edges of the immersed vessel generate surface waves that propagate towards the center of the bell. When the amplitude of the oscillation increases, wave amplitude increases. We study the influence of the angle between successive sides on the wave patterns. Two kinds of vessel have been studied: a triangular and a square prism. The shape of the air/oil meniscus depends on the angle between the sides of the considered prism. As the amplitude of the oscillation is increased, the triple line, which is the contact line between the solid and the air/oil interface, moves up and down. Above a given acceleration that depends on the immersion depth and on the shape vessel, wave goes under the corner edge of the bell. During the oscillation, the wave generates at the edges presents a singularity that leads eventually to a jet and a drop ejection. A drop is ejected at each oscillation. More complicated ejection can be produced with further increase of the amplitude. This is a sample arXiv article illustrating the use of fluid dynamics videos.\end{abstract} 
 

The vertical oscillation of a fluid interface is certainly one of the most prolific field of fluid dynamics since seminal works by M. Faraday \cite{farad}. Above a given acceleration (Faraday threshold), a parametric instability occurs generating half harmonic wave pattern, so called Faraday waves. This state plays a key role in numerous phenomena like wave turbulence \cite{turbu}, droplet manipulation \cite{couder}, quantum-like system \cite{pnas}... The analyze of Faraday waves has been extended to viscous fluids \cite{tuckerman} and to the interface of two fluids \cite{krishna}. In this work, the negative of the Faraday experiment is proposed. Instead of shaking a vessel filled with a liquid in the air, a vessel filled with air is shaken in a liquid. In order to obtain this effect, a vessel is plunged upside down into a large pool of 50 cSt silicone oil (see Fig. 1). This vessel forms a diving bell in the pool and is vertically shaken. The air trapped in the bell induces the vibration of the interface. 
\begin{figure*}
\begin{center}
\includegraphics[width=8cm]{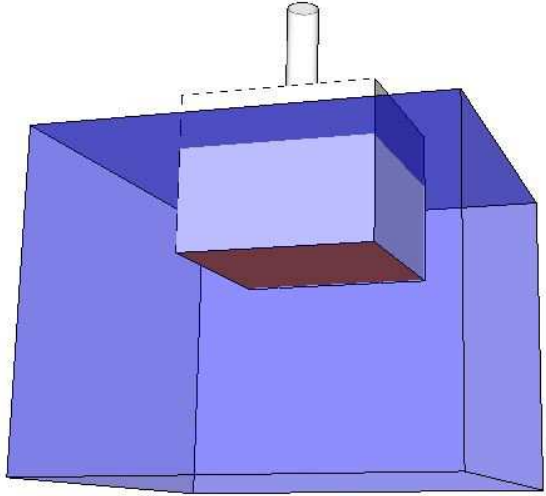}

\end{center}
\caption{Diving bell. The vessel (a square prism in the present case) is vertically shaken at 60 Hz using an electromagnetic shaker. The motion of the submerse interface (in dark) is recorded using a high speed camera.}
\end{figure*}
\begin{figure*}
\begin{center}
\includegraphics[width=12cm]{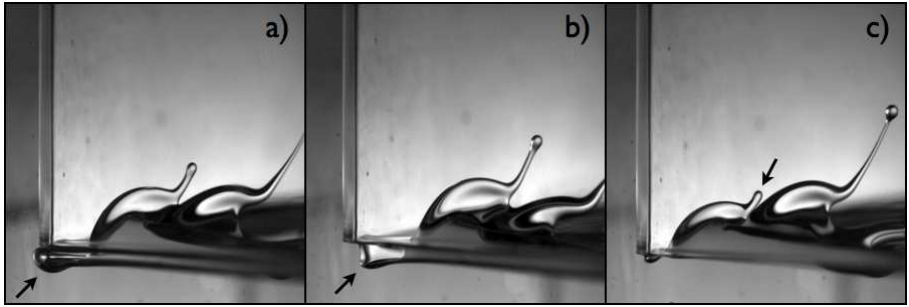}

\end{center}
\caption{Snapshots of a droplet ejection from one corner. (a) Rim formation below the vessel edge, (b) formation of the singularity, and (c) generation of a jet that ejects the droplet.}
\end{figure*}

The triple line that delimits the interface air/oil and the solid in the diving bell depends on the nature and on the shape of the bell. This line is located inside the prism since the hydrostatic pressure compresses the air located in the bell. In this work, two vessel geometries were used: (i) a triangular prism and (ii) a square prism. Both are made in polycarbonate which is rather well wetted by oil. The angle of the corner influences the shape of the meniscus. Indeed, the oil goes higher in an acute angle, namely in the edge of the triangular prism than in the square prism. 

When the prism is shaken the triple line moves up and down following the motion of the bell. When the amplitude of oscillation is sufficient, the triple line can go out of the interior of the prism (Fig.2). More precisely, when the vessel goes upwards, an air roll is formed along the rim of the prism. This air roll is the largest at the edge (Fig.2a). When the vessel goes downwards, the air roll is expelled towards the inner part of the bell. A singularity (cusp-like) occurs at each edge of the prism (Fig.2b) and a jet is observed from the corner towards the center of the prism (Fig.2c). A droplet is eventually ejected (Fig.2d). The acceleration required to obtain this spectacular effect is larger in the triangular prism than in the square prism since the meniscus at edges is higher.

SD thanks FNRS for Þnancial support. Part of this work has been supported by COST P21: physics of droplets (ESF).\\

\end{document}